\documentclass{aa}  
\usepackage{graphicx}
\usepackage{txfonts}
\usepackage{natbib}

\begin{document}
   \title{Revisiting the optical depth of spiral galaxies using the Tully-Fisher B relation}

   \author{E. Kankare
          \inst{1}
          \and 
          M. Hanski
          \inst{1}
          \and
          G. Theureau
          \inst{2,}
          \inst{3}
          \and
          P. Teerikorpi
          \inst{1}
          }

   \institute{Tuorla Observatory, Department of Physics and Astronomy, University of Turku, FI-21500 Piikki\"{o}, Finland\\
              \email{eskank@utu.fi}
              \and
              GEPI/CNRS URA1757, 92195 Meudon Principal Cedex, France
              \and
              LPCE/CNRS UMR6115, 3A Av. de la recherche scientifique, 45071 Orl\'{e}ans Cedex 02, France\\
             }

 
  \abstract
   {}
   {We attempt to determine the optical depth of spiral galaxy disks by a statistical study of new Tully-Fisher data from the ongoing KLUN+ survey, and to clarify the difference between the true and apparent behavior of optical depth.}
   {By utilizing so-called normalized distances, a subsample of the data is identified to be as free from selection effects as possible. For these galaxies, a set of apparent quantities are calculated for face-on positions using the Tully-Fisher diameter and magnitude relations. These values are compared with direct observations to determine the mean value of the parameter $C$ describing the optical depth.}
   {The present study suggests that spiral galaxy disks are relatively optically thin $\tau_{\mathrm{B}} \approx 0.1$, at least in the outermost regions, while they appear in general to be optically thick $\tau_{\mathrm{B}} > 1$ when the apparent magnitude and average surface brightness are studied statistically.}
   {}


   \titlerunning{Optical depth of spiral galaxies}
   \authorrunning{E. Kankare et al.}

   \maketitle
%

\section{Introduction}

  The optical depth of spiral galaxy disks has been debated extensively with widely diverse results. The optical depth is an important parameter to determine because it affects how galaxies appear at different viewing angles. By understanding the inclination dependence of total magnitude, surface brightness, and diameter, inclination corrections for these observed quantities can be determined, and spiral galaxies inclined at different angles and located at different distances can be related. The optical depth of spiral galaxies is at particular interest in supernova surveys because of the need to correct supernova magnitudes \citep{hatano97, hatano98} and supernova rate determinations \citep{riello, botticella} for the effects of extinction. Internal extinction is another important parameter related to selection bias in extragalactic Cepheid distance determinations \citep{paturel05}.

  The classical approach is to derive overall disk extinction from the effects of disk inclination on observables such as magnitude, diameter, and average surface brightness. If there is no dust in the spiral galaxy disks or the dust extinction is negligibly small, the apparent magnitude of the disk appears to be constant, regardless of viewing angle because no material absorbs the radiation and the galactic environment is so diffuse that individual star images, such as point-like sources, do not overlap each other. However, the apparent surface brightness increases as the inclination increases because the area of the ellipse projected by the galaxy image decreases. For the optically thick case, the brightness quantities vary in an opposite way. If there is a lot of dust in the spiral galaxy only the outer layer of the stars can be seen or at least it dominates the brightness observations. The depth of this star layer remains constant inspite of the inclination angle and therefore so does the surface brightness. However, as the inclination increases and the apparent surface area decreases, the apparent total brightness will decrease also.

  Several statistical studies applying the classical or similar methods have claimed that spiral galaxy disks are optically thick at least in their inner parts. These include \citet{valentijn}, \citet{burstein} including distance information, \citet{giovanelli} studying surface brightness shifts, \citet{peletier} using  multiple wavebands, \citet{bottinelli95} utilizing Tully-Fisher relation (hereafter TF relation) with normalized distances, \citet{tully} with multiple wavebands and volume limited galaxy groups, \citet{shao} studying the luminosity function, and \citet{driver} using a inclination relation model for the galaxy components. 

  Although the studies cited above and related papers imply that spiral galaxy disks should be considered to be optically thick, conflicting results have been obtained from statistical studies. From diameters, \citet{huizinga} inferred that galaxy disks are optically neither completely thick nor thin. \citet{davies} concluded that it is impossible to determine optical depth with statistical studies because of selection effects. \citet{jones} criticized the way in which galaxy samples had been used to obtain a volume-limited statistical sample.

  Statistical studies of the optical depth of spiral galaxies with samples of several hundreds or thousands of galaxies involve multiple selection effects, therefore inspecting only a few galaxies closely, represents a completely different way to approach the problem. Investigations limited to edge-on galaxies usually led to the conclusion that the galaxies are optically thin. These included the studies of \citet{bosma} and \citet{byun93}, which compared rotation curves, and \citet{xilouris97,xilouris98,xilouris99}, which applied various exponential fits to observations of stars and dust.
  
  For face-on spirals, a more direct method has been used: one observes background objects through the target galaxy. These studies suggest that the galaxies are optically rather thin, with optically thicker spiral arms and thinner interarms. A simpler form of this method is to observe either galaxy pairs, as achieved by \citet{andredakis} and \citet{berlind}, or multiple pairs, as observed by \citet{white00} and \citet{domingue}. Another method of observing background objects through spiral galaxies is the so-called synthetic field method introduced by \citet{gonzalez}, who studied the amount of distant galaxies observable through the parts of spiral galaxy disks compared with reference fields. \citet{holwerda05a} improved the method with automation, and a study of several galaxies was presented by \citet{holwerda05b}. 

  In contrast, the Milky Way disk is considered to be optically thin. At the Galactocentric solar distance, \citet{koppen} inferred a value $A_{\mathrm{V}} = 0.1$ for the face-on absorption. However, even at high latitudes there appears to be directions in which there is higher local extinction e.g. \citet{berdyugin}.

  In this study, we aim to determine the optical depth of the spiral galaxy disks and inclination corrections for diameter and magnitude in the new TF data sample obtained from the still ongoing KLUN+ (Kinematics of the Local Universe\footnote{http://klun.obs-nancay.fr/KLUN+/page1.html}) survey. The structure of the paper is the following. In Sect. 2, the method used for the old KLUN sample by \citet{bottinelli95} is presented. In Sect. 3, the method is applied to a new sample and the results are given. In Sect. 4, an additional test of the method is presented using a large galaxy sample obtained from the extragalactic database HyperLeda\footnote{http://leda.univ-lyon1.fr/}. In Sect. 5, the difference between true and apparent optical depth is discussed and in Sect. 6, a summary is given. 


\section{TF relation in the classical approach}

\subsection{Influence of selection effects}

  Almost every observed galaxy sample is limited by either diameter or magnitude or both, due to the limitations of the observational program and the telescope adopted. Therefore, it is useful to summarize briefly how the classical approach, utilizing diameters and magnitudes versus inclination, is affected by the inevitable limits in angular size and/or apparent magnitudes. It is helpful to consider first the idealized case of a galaxy cluster consisting of a single type of galaxy, with a small scatter in linear diameter (or absolute magnitude). This can be easily generalized to field galaxies, which are localized into different bins of radial velocity, and therefore different ``clusters'', at different distances. If the sample is diameter limited and galaxies are opaque, the isophotal diameter does not depend on inclination: the sample is therefore representative of all inclinations and the diameter limit does not affect the (zero) slope. Classes of smaller galaxies just remain unobserved. 

\begin{figure}
   \centering
   \includegraphics[angle=-90,width=88mm]{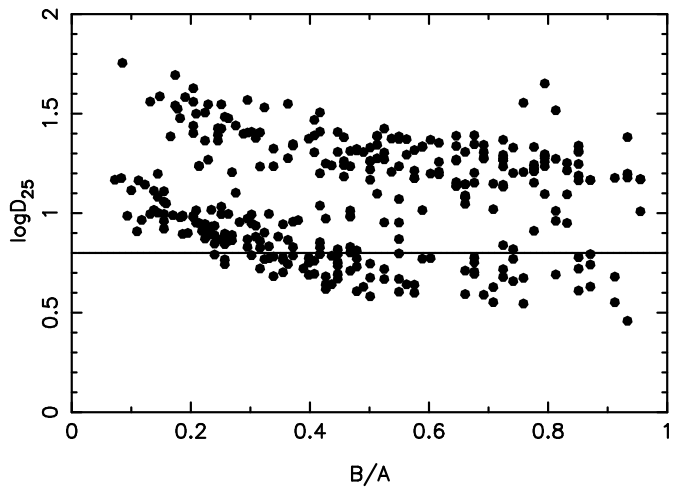}
   \caption{Samples of spiral galaxies from the HyperLeda database separated into two groups with apparent magnitude $13.0 < m < 13.2$ above and $16.5 < m < 16.7$ below. If limited by angular diameter, i.e. line $\log D_{\mathrm{25}} = 0.8$, the sample would lack face-on (axis ratio $B/A \rightarrow 1$) galaxies in the non-opaque case, and the slope would be artificially too shallow.}
   \label{SelA}
   \end{figure}

  In non-opaque case, the diameter increases with inclination; the low inclination part of the sample is then less complete, which produces too shallow a slope in the $\log D$ versus $B/A$ diagram, and therefore too high an opacity. The axis relation $R=A/B$ is used if logarithmic scaling is needed where $B$ is the minor axis and $A$ the major axis of the galaxy. The sample completeness example constructed from the HyperLeda database is shown in Fig.~\ref{SelA}.

\begin{figure}
   \centering
   \includegraphics[angle=-90,width=88mm]{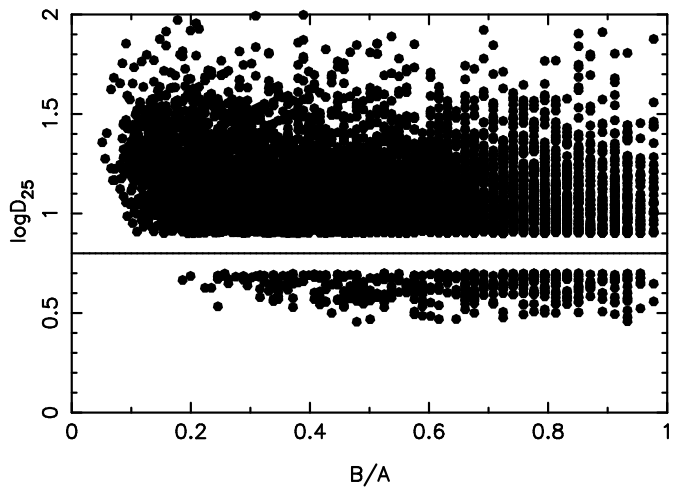}
   \caption{Samples of spiral galaxies from the HyperLeda database separated into two groups with apparent diameter $\log D_{\mathrm{25}} > 0.90$ and $\log D_{\mathrm{25}} < 0.70$. If limited with diameter, i.e. line $\log D_{\mathrm{25}} = 0.8$, the sample would be complete in a non-opaque case.}
   \label{SelB}
   \end{figure}

  In a magnitude-limited sample, the selection effect is opposite. In the opaque case, the magnitude depends on inclination (edge-on galaxies are fainter), so the high inclination part of the sample is less complete, which again produces too shallow slope in the $m$ versus $B/A$ diagram, and in this case artificially low opaqueness. In fully non-opaque cases, all inclinations are equally completely represented (see for comparison Fig.~\ref{SelB}, for the trivial case).

  To avoid these effects, one may use the TF relation for magnitude or diameter. One constructs the Hubble parameter $H$ versus normalized kinematic distance diagram to find the unbiased plateau, where the sample is unaffected by the diameter (or magnitude) limit and should be representative of all inclinations \citep{bottinelli95}.

\subsection{The relevant quantities}

  For every spiral galaxy the apparent Hubble constant is calculated by an infall model around the Virgo cluster, for which
  \begin{equation} \label{eq:MetA}
      H_{\mathrm{0}} = \frac{(V_{\mathrm{0Vir}} + v_{\mathrm{0}})d_{\mathrm{kin}}}{r} \,,
  \end{equation} 
where the distance $r$ originates in $M = m - 25 - 5 \log r$, in which the absolute magnitude $M$ is evaluated by the TF relation 
  \begin{equation} \label{eq:MetB}
      M = a (\log V_{\mathrm{m}} - 2.2) + b \,,
  \end{equation} 
  where $a$ and $b$ are the TF parameters. Alternatively, one may also use the diameter TF relation. In what follows, the parameters of the diameter relation are denoted by $a'$ and $b'$.

  A short review of the equations derived and the galaxy model applied to the spiral galaxies by \citet{bottinelli95} is given. The inclination dependence of apparent magnitude given in the form $m = m_{\mathrm{0}} + A(R)$ is  
  \begin{equation} \label{eq:A}
      m = m_{\mathrm{0}} - 2.5 \log (k+(1-k)R^{2C(1+{0.2/K})-1}) \,.
   \end{equation}
We adopted the constants $K = 0.1$ from \citet{fouque}, when the morphological type was $T \geq 2$, and $k = 0.16$ from \citet{simien} as a mean value when $2 \leq T \leq 7$. When the apparent surface brightness is given by 
  \begin{equation} \label{eq:myy}
      \mu = m + 5 \log D_{\mathrm{25}} \,,
   \end{equation}
the inclination dependence of apparent surface brightness is easily derived in the form $\mu = \mu_{\mathrm{0}} + B(R)$ as
  \begin{equation} \label{eq:B}
      \mu = \mu_{\mathrm{0}} - 2.5 \log (kR^{-2C}+(1-k)R^{(0.4C/K)-1}) \,.
   \end{equation}

\subsection{Results by \citet{bottinelli95}}

\begin{figure}
   \centering
   \includegraphics[angle=-90,width=88mm]{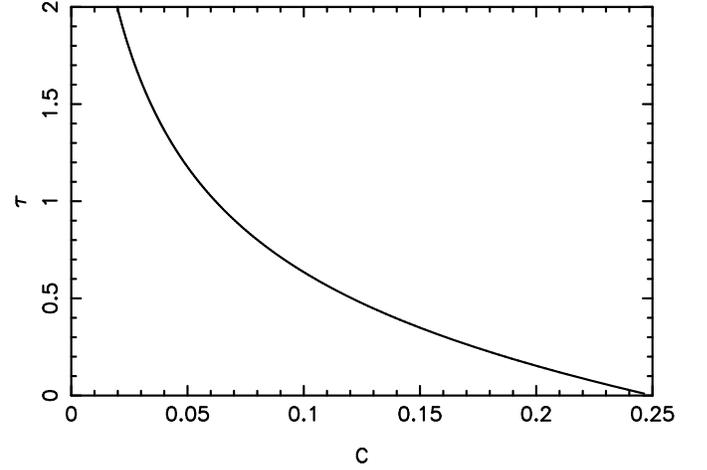}
   \caption{Dependence of optical depth $\tau $ on the parameter $C$ used in our method.}
   \label{Tau}
   \end{figure}

  The initial sample of \citet{bottinelli95} included 4535 spiral galaxies. Using the diameter TF relation, an unbiased subsample was identified. This subsample consisted of 332 galaxies. The value obtained for the subsample $C=0.07 \pm 0.05$ corresponds to $\tau \approx 0.9$ (see Fig.~\ref{Tau} for the dependence of $\tau $ on $C$). However, it should be noted that the optical depth of the disk is expected to decrease as the radius increases. Therefore, parts of the inner disk may have $\tau > 1$. The limitation of the plateau subsample did not change the result significantly. For the unlimited ``biased'' sample, the result $C=0.05 \pm 0.02$ was obtained. 

\section{New sample}

\subsection{Basic sample}

The HI data used in this study were acquired as part of the KLUN project, based on observations with the Nan\c{c}ay radio telescope during the years 1990-2006, for the purpose of distance and peculiar velocity measurements. Approximately 5000 HI profiles\footnote{Data tables and HI-profiles and corresponding comments are available in electronic form
at the CDS via anonymous ftp to cdsarc.u-strasbg.fr (130.79.128.5) or via http://cdsweb.u-strasbg.fr/Abstract.html. They are also provided through Simple Spectra 
Access and partly available at NED (http://nedwww.ipac.caltech.edu/).} have been gathered at Nan\c{c}ay and corresponding data published in the KLUN series papers \citep{theureau98, theureau05, theureau07, theureau08, paturel03b}.

All KLUN HI spectra acquired by the Nan\c{c}ay radio telescope antenna were reviewed and assigned a quality code according to the shape of their 21-cm line profile. This study \citep[][and references above]{guillard} has enabled us to identify efficiently a large number of TF outliers due to morphological type mismatch or HI-confusion in the elongated beam of Nan\c{c}ay.

The photometric input catalogue was compiled from the Hyperleda extragalactic database:
\begin{itemize}
\item[-] Accurate coordinates (of typical accuracy lower in value than 2 arcsec), diameter, axis ratio, and position angle were extracted from the analysis of the Digitalized Sky Survey (DSS); Diameters and axis ratios were corrected to the RC2 system at the limiting surface brightness of 25 mag arcsec$^{-2}$ \citep{paturel03a}.
\item[-] \textit{B}-band magnitudes were extracted both from a compilation of \textit{B}-band photometry available in the literature and, in a homogeneous way, directly from the DSS \citep{paturel00}. Each magnitude is the result of a weighted mean over all references, after correction to the RC3 system \citep[see][]{paturel97}.
\item[-] To complete the Galactic extinction correction, $a_{\rm g}$, we used values adopted from \citet{schlegel}.
\end{itemize}

The morphological classification in HyperLeda is based on visual inspection. As for the other parameters, each type is obtained as a weighted mean of the $T$ parameter \citep{vaucouleurs91} over existing references in the literature. This number $T$ is between -5 and -2 for ellipticals, between -2 and 1 for lenticulars, 1 to 8 for spiral galaxies, and 9 and 10 for irregular. The DSS image of each KLUN target was inspected by eye before HI observation, to confirm its spiral galaxy nature and determine whether a bar was present. The Sa type spiral galaxies were excluded from the sample because of the possibility that the bulge of the galaxy may dominate the light detected during observations. Galaxies close to the Galactic plane $ \vert b \vert \leq 15\,^{\circ}$ with large galactic extinction were excluded.

\begin{figure}
   \centering
   \includegraphics[angle=-90,width=88mm]{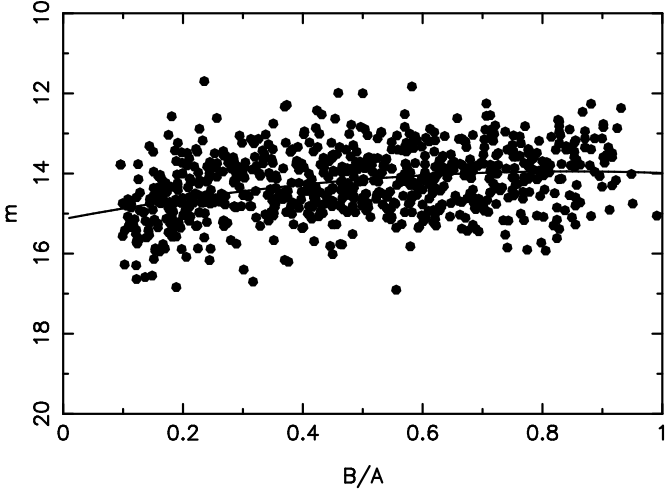}
   \caption{Inclination dependence of magnitude for the initial sample of 772 galaxies with the mean trend plotted.}
   \label{FigMag}
   \end{figure}

\begin{figure}
   \centering
   \includegraphics[angle=-90,width=88mm]{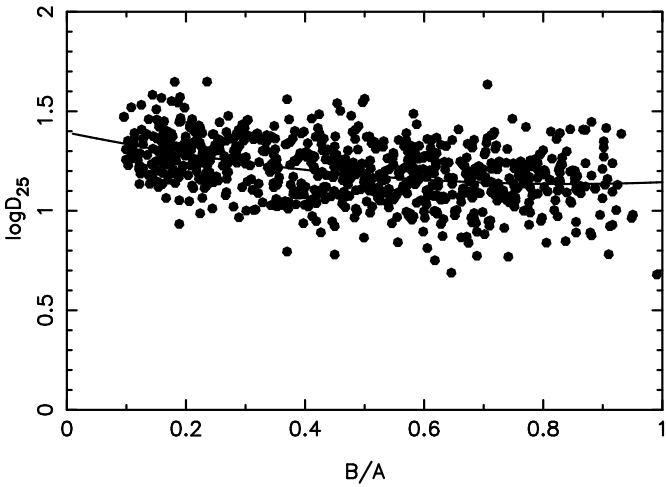}
   \caption{Inclination dependence of diameter for the initial sample of 772 galaxies with the mean trend plotted.}
   \label{FigDiam}
   \end{figure}

The adopted method of normalized distances utilizes the Virgo cluster infall model, with parameters $RA_{\mathrm{Vir}}=12.5\,\mathrm{h}$, $\delta_{\mathrm{Vir}}=12\,^{\circ}$, and $V_{\mathrm{0Vir}}=980\,\mathrm{km/s}$. For the infall speed of the Local group, a value of $v_{\mathrm{0}}=150\,\mathrm{km/s}$ is used as in \citet{bottinelli95}. Galaxies at small angular distance from the Virgo cluster $\theta < 30\,^{\circ}$ and Virgo members are excluded so that unambiguous values of $d_{\mathrm{kin}} = r/r_{\mathrm{Vir}}$ can be determined from the kinematic velocity field model \citep{bottinelli95}. Here $r$ is the distance of the sample galaxy and $r_{\mathrm{Vir}}$ the distance of the Virgo cluster so $d_{\mathrm{kin}}$ and later on $d_{\mathrm{n}}$ are dimensionless quantities. All of these restrictions further limit the size of the sample to 772 individual galaxies. From here on, this is referred to as the initial sample unless otherwise specified. The inclination dependence of magnitude and diameter for the initial sample is presented in Fig.~\ref{FigMag} and \ref{FigDiam}, respectively.

In Fig. \ref{FigMag}, it can be seen that the magnitude distribution of the brightest and closest galaxies is rather smooth. Regardless of how the distant galaxies are selected, the distribution of galaxies with the lowest $m$ values in the sample differs significantly e.g., from the ESO-LV \citep{lauberts} sample in which the brightest edge-on galaxies are on average $1 - 1.5\,\mathrm{mag}$ fainter than the brightest face-on galaxies. This can be seen in several opacity studies such as \citet{huizinga} and \citet{burstein}.

The initial sample can be taken to be complete for a diameter limit of $\log D_{\mathrm{25}}=1.1$ and magnitude limit of $m_{\mathrm{T}} \approx 15$. The sample therefore contains more distant galaxies than the data used by \citet{bottinelli95}. The sample is not limited by magnitude, although, galaxies with $\log D_{\mathrm{25}}<1.1$ are restricted from the sample. The diameter limit is also used in Eq. (\ref{eq:dn}). From the diameter distribution in Fig. \ref{FigDiam}, it can be seen that the sample lacks faint edge-on galaxies. This was also noted by \citet{bottinelli95}, who attributed the lack to a systematic error in which the inclinations of small galaxies were measured to be too small. The same problem was discussed by \citet{giovanelli}, who claimed that it was a problem caused by the resolution limit of observations. Whatever the source of the problem is, we use the same limitations as \citet{bottinelli95}, who excluded galaxies with $\log (R)>0.8$ $(B/A<0.16)$. Face-on galaxies with $\log (R)<0.07$ $(B/A>0.85)$ are excluded because they lack accurately measured rotation speed and are of little use to the TF relation. 

This limits the sample to 495 galaxies. This is referred to as the basic sample unless otherwise specified. For the basic sample, initial TF relation parameters were determined by fitting slopes to a relation between linear diameter and the maximum rotational velocity with a value of $a'=1.1$, and to a relation between absolute magnitude and rotational velocity with values $a=-6.3$ and $b=-19.52$. A zero point $b'$ for the diameter is insignificant in the adopted method because the face-on diameters determined by the TF relation are scaled to the observations with a fitting constant as seen in Eq. (\ref{eq:S}).

\subsection{The diameter limited subsample}

\begin{figure}
   \centering
   \includegraphics[angle=-90,width=88mm]{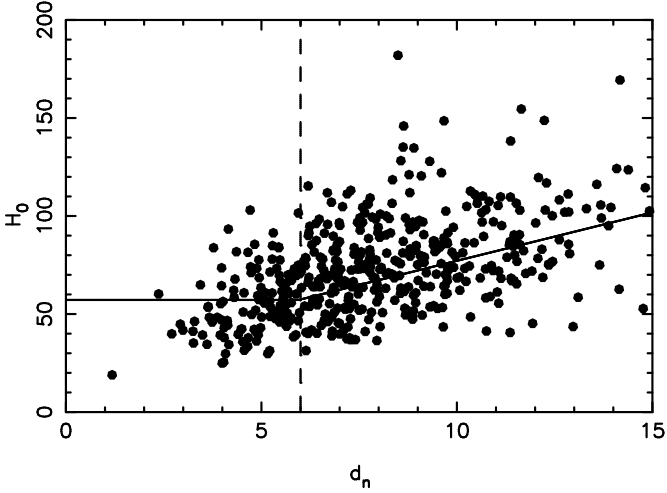}
   \caption{Hubble constant versus the normalized distance for the basic sample. The curve represents the constant value of the Hubble parameter for the subsample when $d_{\mathrm{n}} \leq 6$ and the increase of the parameter with a linear fit when $d_{\mathrm{n}} > 6$.}
   \label{FigHub}
   \end{figure}

The basis of the normalized distance method is to extract at each distance, a subsample free of selection effects, i.e. a subsample for which the average TF distance is its true mean distance. Normalized distances for the basic sample are derived from the kinematic distances by an equation from \citet{bottinelli95}
  \begin{equation} \label{eq:dn}
      d_{\mathrm{n}} = d_{\mathrm{kin}}10^{a'(2.7-\log{V_{\mathrm{m}}})} \cdot 10^{-C \log{(A/B)}} \cdot 10^{0.094a_{\mathrm{g}}} \cdot 10^{\log{D_{\mathrm{l}}}-1.3} \,,
   \end{equation}
where $\log D_{\mathrm{l}}$ is the diameter limit for the sample determined above and $a_{\mathrm{g}}$ is the galactic extinction. The optical depth parameter has an initial value of $C=0$. The average value of the Hubble constant is calculated for galaxies at the same normalized distance $d_{\mathrm{n}}$. As expected from the selection effect that removes absolutely small galaxies, the value of $H_{\mathrm{0}}$ is seen to increase at large normalized distances. This effect is seen in Fig.~\ref{FigHub} at $d_{\mathrm{n}}>6$. With smaller values, the Hubble constant has an average value of $H_{\mathrm{0}}\approx 60\,\mathrm{km\,s^{-1}\,Mpc^{-1}}$ forming an unbiased sample consisting of 122 spiral galaxies. This is referred to as the diameter-limited (DL) subsample unless otherwise specified. The diameter TF parameter with a value $a'=1.2$ was determined for the DL subsample.

\begin{figure}
   \centering
   \includegraphics[angle=-90,width=88mm]{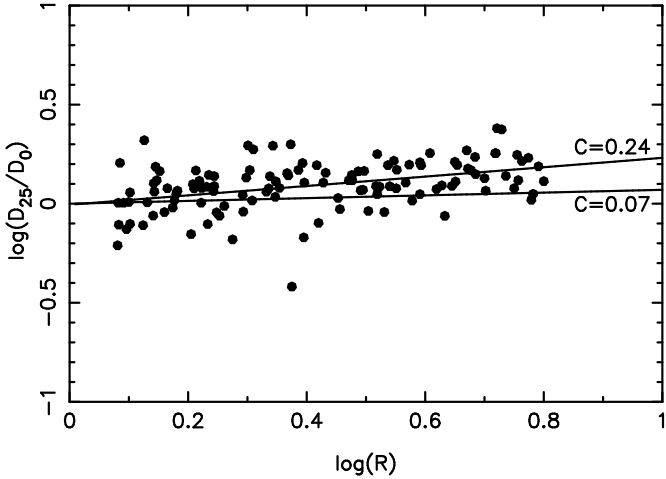}
   \caption{$\log(D_{\mathrm{25}}/D_{\mathrm{0}})$ versus $\log(R)$ for the DL subsample. The upper line is a fit for the data with a slope $C=0.24$. The lower line represents the result of \citet{bottinelli95} with a slope $C=0.07$.}
   \label{FigC1}
   \end{figure}

  For the DL subsample, the apparent reference diameters that the galaxies should have if seen face-on are determined with the TF relation 
  \begin{equation} \label{eq:S}
      \log D_{\mathrm{0}} = - \log d_{\mathrm{kin}} + a'\log{V_{\mathrm{m}}} + cst \,.
   \end{equation}
The parameter $C$ that provides the average optical depth of the DL subsample is determined from the equation
  \begin{equation} \label{eq:D}
      \log D_{\mathrm{25}} = \log D_{\mathrm{0}} + C \log (R) \,.
   \end{equation}
The inclination dependence of $\log(D_{\mathrm{25}}/D_{\mathrm{0}})$ is presented in Fig.~\ref{FigC1}. The line fit to the data is completed with a least squares method and has a slope value of $C=0.24 \pm 0.05$. It should be mentioned that for the method used here, of all observed values the accuracy of the apparent diameter has the most significant effect on the result.

The value obtained for parameter $C$ corresponds in terms of optical depth to $\tau \approx 0.1$ which would imply that spiral galaxies are extremely optically thin. However, as expressed in Sect. 2.2,  it is expected that the optical depth decreases as the radius increases. The result $C=0.24$ for the DL subsample is true only for both the diameter and the outer rim of the disk, and does not provide any information about the optical depth in the inner parts of the disk or even about the average optical depth for the entire disk. This is because the method relies on diameter observations that are affected mostly by the optical depth in the outer parts of the galaxy disk. 

\subsection{The magnitude limited subsample}

\begin{figure}
   \centering
   \includegraphics[angle=-90,width=88mm]{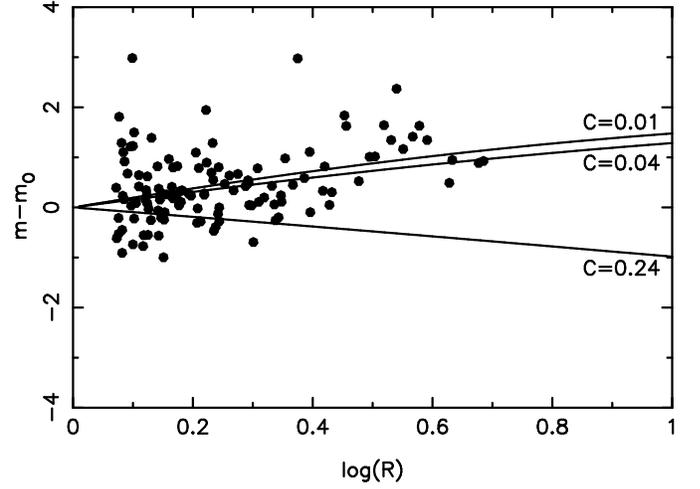}
   \caption{Inclination dependence of magnitude difference $m-m_{\mathrm{0}}$. The curves from top to bottom represent the function $A(R)$ with $C=0.01$ for these data, $C=0.04$ by \citet{bottinelli95}, and $C=0.24$ for the DL subsample, respectively.}
   \label{FigA}
   \end{figure}

The behavior of the brightness quantities and derived parameter value $C$ can be tested with the functions presented in Eqs. (\ref{eq:A}) and (\ref{eq:B}). Since the equations adopt brightness observations for the entire galaxy disk projected on the sky, there is a need for another subsample. The initial sample of 772 galaxies is now limited as before but by the magnitude completeness limit instead of the diameter limit providing another basic sample of 574 galaxies with TF parameter values of $a=-5.7$ and $b=-19.67$. The normalized distance is now determined for magnitude purposes by equation 
  \begin{equation} \label{eq:dn2}
      d_{\mathrm{n}} = d_{\mathrm{kin}}10^{0.2a(\log{V_{\mathrm{m}}}-2.7)} \cdot 10^{0.2A(R)} \cdot 10^{0.2a_{\mathrm{g}}} \cdot 10^{-0.2(m_{\mathrm{T}}-13)} \,,
   \end{equation}
where inclination correction has an initial value of $A(R)=2.5 \log R$, which corresponds to $C=0$ and an opaque disk. The behavior of the Hubble constant versus the normalized distance is similar regardless of the way in which the distance is determined from the kinematic distance, and again galaxies with the normalized distance $d_{\mathrm{n}}>6$ are excluded providing another subsample of 124 galaxies with essential parameter values of $a=-6.5$ and $b=-19.19$. For each galaxy, the apparent reference magnitude if seen face-on is calculated using the TF relation
  \begin{equation} \label{eq:m0}
      m_{\mathrm{0}} = 5 \log d_{\mathrm{kin}} + a \log V_{\mathrm{m}} + cst \,.
   \end{equation}
The function $A(R)$ and the ML subsample are presented in Fig.~\ref{FigA}. The best-fit function to the data provides as low a value as $C=0.01 \pm 0.05$, corresponding to $\tau > 1$, which agrees fairly well with the result of $C=0.04$ from \citet{bottinelli95}. 

\begin{figure}
   \centering
   \includegraphics[angle=-90,width=88mm]{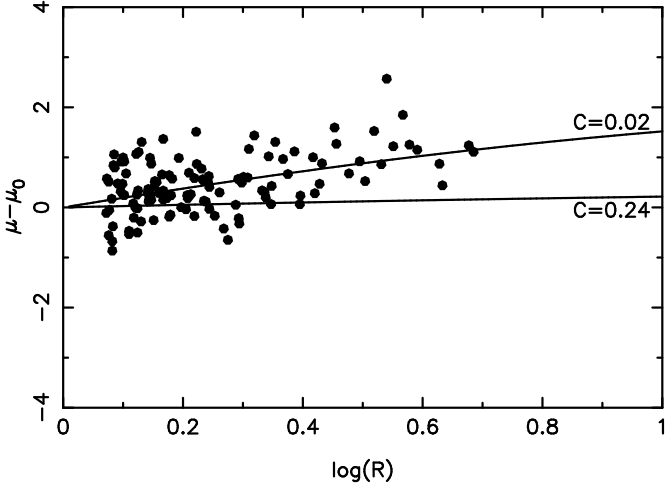}
   \caption{Inclination dependence of surface brightness difference $\mu - \mu _{\mathrm{0}}$.  The curves represent the function $B(R)$. The upper curve is a fit for the data with a value $C=0.02$. The lower curve represents the result from the DL subsample with a value $C=0.24$.}
   \label{FigB}
   \end{figure}

Similarly the function $B(R)$ is presented in Fig.~\ref{FigB}. The apparent face-on surface brightness $\mu _{\mathrm{0}}$ that the galaxy should have is derived from Eqs. (\ref{eq:myy}) and (\ref{eq:m0}). The best-fit description of the data is $C=0.02 \pm 0.07$ (corresponding again to $\tau > 1$). Since the total brightness difference between face-on and edge-on galaxies is large, it dominates the surface brightness difference regardless of the increase in apparent diameter due to the optically thin outer galaxy disk. 

\subsection{Other subsamples}

\begin{figure}
   \centering
   \includegraphics[angle=-90,width=88mm]{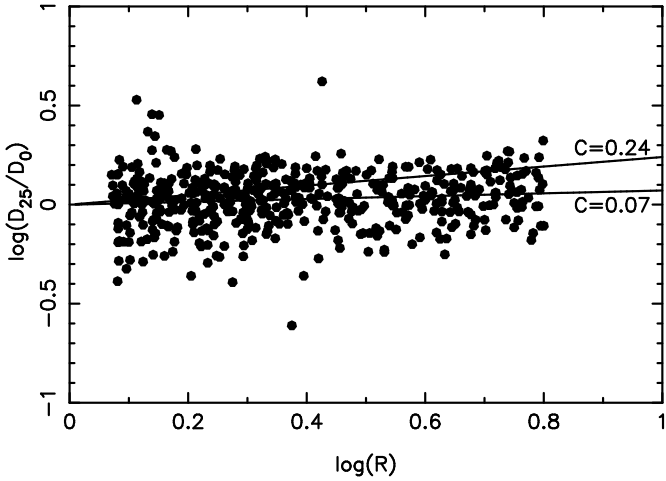}
   \caption{$\log(D_{\mathrm{25}}/D_{\mathrm{0}})$ versus $\log(R)$ for the basic sample. The upper line represents the result from the DL subsample with a slope $C=0.24$. The lower line is a fit for the data with a slope $C=0.07$.}
   \label{FigC2}
   \end{figure}

Additional information can be obtained by restricting the sample in different ways. The optical depth of the basic sample can be determined by applying limits to the diameter as achieved for the DL subsample above. The method provides a value $C=0.07 \pm 0.03$, which agrees with the results of \citet{bottinelli95}. However, the difference between the DL subsample and the basic sample for the new data is not unexpected. The result derived for the DL subsample implied that spiral galaxy disks are optically thin at least in terms of the inclination dependence of the diameter; this would imply that the basic sample limited to certain values of diameter is incomplete and lacks faint, low-inclination-angle, galaxies, as stated in Sect. 2.1. A graphical presentation is seen in Fig.~\ref{FigC2}.

\begin{figure}
   \centering
   \includegraphics[angle=-90,width=88mm]{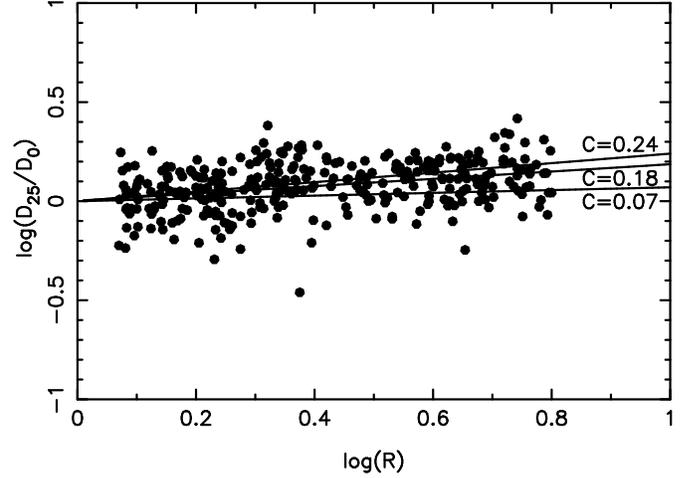}
   \caption{$\log(D_{\mathrm{25}}/D_{\mathrm{0}})$ versus $\log(R)$ for the DL subsample including low-quality data. The lines from top to bottom represent $C=0.24$ for the basic DL subsample, $C=0.18$ for these data, and $C=0.07$ by \citet{bottinelli95}, respectively.}
   \label{FigC3}
   \end{figure}

We note that \citet{bottinelli95} considered galaxies with HI profile observations that had only one peak or a low S/N ratio. To avoid a possible source of error only galaxies with high quality HI profile data were included in the samples used above to determine the optical depth. For comparison the data samples could be expanded to include also low-quality observations. This increases the diameter-limited basic sample to 835 galaxies with TF parameters of $a'=0.9$, $a=-4.7$, and $b=-19.67$. When limited by the normalized distance, a subsample of 332 galaxies is obtained with a TF parameter of $a'=1.1$. The method then provides a slope value $C=0.18 \pm 0.03$ for the opacity parameter with the fit presented in Fig.~\ref{FigC3}. The result corresponding to a value $\tau \approx 0.2$ is relatively close to the value obtained for the DL subsample; it therefore appears unable to explain the differences between our results and those of \citet{bottinelli95}.

\begin{figure}
   \centering
   \includegraphics[angle=-90,width=88mm]{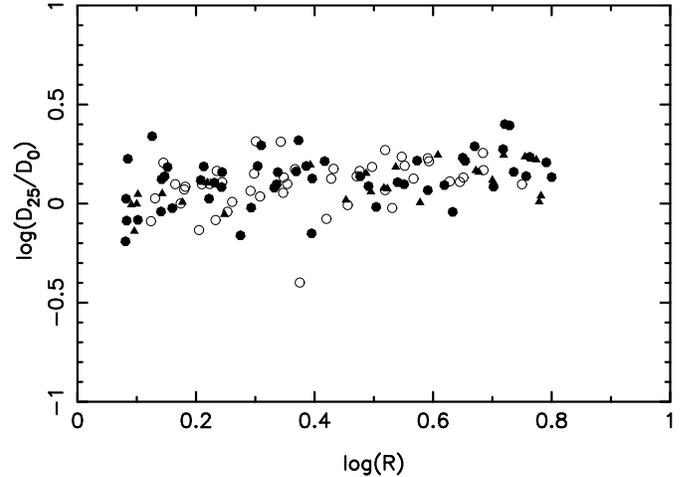}
   \caption{$\log(D_{\mathrm{25}}/D_{\mathrm{0}})$ versus $\log(R)$ for the DL subsample with different morphological types separated as Sab-Sb $(\circ)$ galaxies, Sbc-Sc $(\bullet)$ galaxies, and Scd-Sd $(\blacktriangle)$ galaxies. See text for details.}
   \label{FigC4}
   \end{figure}

Finally, the DL subsample, as can be seen in Fig.~\ref{FigC4}, is divided into closer morphological types. The dispersions and results for the categories are: 46 type $2 \leq T \leq 3.5$ Sab-Sb galaxies with a value $C=0.24 \pm 0.10$, 51 type $3.5 < T < 5.5$ Sbc-Sc galaxies with a value $C=0.23 \pm 0.08$, and 25 type $5.5 \leq T \leq 7$ with a value $C=0.25 \pm 0.07$. There appears to be no significant differences between morphological classes, all types being optically very thin at least in their outer parts.

\section{HyperLeda sample}

\begin{figure}
   \centering
   \includegraphics[angle=-90,width=88mm]{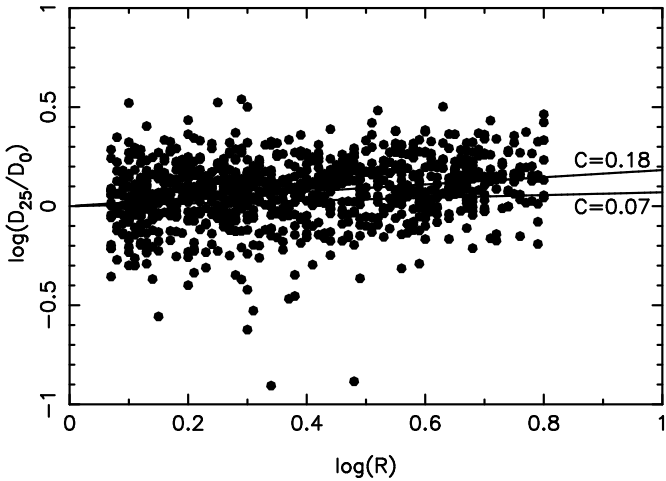}
   \caption{$\log(D_{\mathrm{25}}/D_{\mathrm{0}})$ vs. $\log(R)$ for the HyperLeda sample. The upper line is a fit for the data with a slope $C=0.18$. The lower line represents the result of \citet{bottinelli95} with a slope $C=0.07$.}
   \label{FigL}
   \end{figure}

Because the data sample used in Sect. 3 was small, a more statistical study was completed for a far larger galaxy sample that included HI observations obtained from the HyperLeda. No restrictions were placed on the quality of the radio observations. The sample includes 9580 spiral galaxies from which galaxies with a small angular distance to the galactic plane and the Virgo cluster were excluded, such that the sample was similar to the initial sample in Sect. 3.1. Using similar limitations, a selection-effect-free subsample of 989 galaxies with $d_{\mathrm{n}}<5$ was obtained with the TF relation. With the method of normalized distances and Eqs. (\ref{eq:MetA})-(\ref{eq:MetB}) and  (\ref{eq:dn})-(\ref{eq:D}), a slope value of $C=0.18 \pm 0.03$ was obtained, which was equivalent to $\tau \approx 0.2$ for the $\log (D_{\mathrm{25}}/D_{\mathrm{0}})$ versus $\log R$ relation (see Fig.~\ref{FigL}). This result correlates well with the results for the subsample in Sect. 3.2. 

  For a magnitude-limited subsample, a sample of 345 spiral galaxies was obtained from the HyperLeda data with the normalized distance $d_{\mathrm{n}}<1.5$. The method represented by Eqs. (\ref{eq:A})-(\ref{eq:B}) and  (\ref{eq:dn2})-(\ref{eq:m0}) corresponds to a parameter value $C=0.02 \pm 0.05$ from the total magnitude relation and $C=0.00 \pm 0.06$ from the surface brightness relation with both corresponding to $\tau > 1$. This again agrees well with the results in Sect. 3.2. All results obtained from different samples for the optical depth parameter $C$ with error limits, used TF parameters, normalized distance limits, and sample sizes are presented in Table \ref{table:1}.

\begin{table*}
\begin{minipage}[t]{\columnwidth}
\caption{Limiting normalized distances, number of galaxies, used TF parameters, and results for different samples.}              
\label{table:1}      
\centering                                      
\renewcommand{\footnoterule}{}  
\begin{tabular}{ c c c c c c c c c c}          
\hline\hline                        
$N$ & Name & $\leqslant d_{\mathrm{n}}$ & Size & $a'$ & $a$ & $b$ & $C$ & $C_{\mathrm{err}}$ & $\tau$ \\    
\hline                                   
    1 & DL subsample & 6.0 & 122 & 1.2 & $-7.5$ & $-18.92$ & 0.24 & 0.05 & 0.1 \\      
    2 & ML subsample ($m$) & 6.0 & 124 & - & $-6.5$ & $-19.19$ & 0.01 & 0.05 & $>1$ \\
    3 & ML subsample ($\mu$) & 6.0 &  124 & - & $-6.5$ & $-19.19$ & 0.02 & 0.07 & $>1$ \\
    4 & DL basic sample & - & 495 & 1.1 & $-6.3$ & $-19.52$ & 0.07 & 0.03 & 0.9 \\
    5 & DL subsample$^{\mathrm{a}}$ & 6.0 & 332 & 1.1 & $-6.5$ & $-19.12$ & 0.18 & 0.03 & 0.2 \\
    6 & DL subsample$^{\mathrm{b}}$ & 5.0 & 989 & 1.0 & $-5.7$ & $-19.53$ & 0.18 & 0.03 & 0.2 \\
    7 & ML subsample$^{\mathrm{b}}$ ($m$) & 1.5 & 345 & - & $-5.1$ & $-19.13$ & 0.02 & 0.05 & $>1$ \\
    8 & ML subsample$^{\mathrm{b}}$ ($\mu$) & 1.5 & 345 & - & $-5.1$ & $-19.13$ & 0.00 & 0.06 & $>1$ \\
\hline      
\end{tabular}
\begin{list}{}{}
\item[$^{\mathrm{a}}$] Same as in $N=1$, now including low-quality data.
\item[$^{\mathrm{b}}$] HI data from the HyperLeda.
\end{list}
\end{minipage}
\end{table*}


\section{Discussion}

\subsection{Effects that increase apparent optical depth}

It appears essential to divide the true optical depths of spiral galaxy disks from those values inferred observationally from statistical arguments. As individual low-inclination-angle galaxies have been studied extensively utilizing different techniques, such as observing galaxy pairs or using the synthetic field method, it appears that face-on galaxy disks are quite optically thin. However, statistical results infer the presence of thicker disks. There are at least three possible reasons for this disagreement because the statistical studies concentrate usually on the inclination dependence of apparent brightness quantities and the brightness difference between face-on and edge-on galaxies in a classical way: light scattered by dust, the geometric distribution of the absorbing dust, and the uncertainty in diameter.

Light scattering can affect the apparent optical depth. \citet{byun94}, \citet{bartolomeo}, \citet{xilouris97}, and \citet{baes} studied several models of light scattering in galaxy disks. Their models indicated that even if spiral galaxies are relatively optically thin, part of the light emitted by the galaxy itself scatters from the thin dust and will increase the apparent luminosity of the low inclination galaxies. When individual galaxies are studied, this has no effect, for example, because background objects are observed through the galaxy under study. The scattering from the light emitted from the background objects can also be neglected as demonstrated by studies such as \citet{berlind} and \citet{white00}. However, if optical depth in galaxies is studied statistically, internal scattering will cause apparent brightness to have an inclination dependence, which may be interpreted classically as evidence for optically thick galaxy disks. 

The second effect is the geometric distribution of dust. Close studies (see Sect. 1) concluded that spiral galaxies are thin not only over the entire disk, but are divided into optically thicker spiral arms and thinner interarms. This type of absorbing medium structure implies that a classical sandwich model is too simplistic for describing spiral galaxies. This is, because the ratio between the projected areas of the thicker spiral arms and the galaxy disk increases as the inclination angle increases. This will again increase the apparent brightness of face-on galaxies relatively and cause disks to behave as if they were optically thicker if studied statistically. However, it should be noted that this effect is minimized by the fact that optically thicker spiral arms are also the brightest parts of the disks and therefore the dominant components in any surface or total brightness studies. 

The third possible effect for poor quality data is the error in measured diameters. \citet{giovanelli} compared scale lengths and surface brightness diameters of observed galaxies with standard table values concluding a decrease in the observable surface brightness levels. Therefore, visual determination of the apparent diameter without using the apparent surface brightness levels would underestimate the size (and also therefore the apparent total brightness) of the edge-on galaxies.

These effects increase the apparent total brightness difference between face-on and edge-on galaxies and therefore have an impact on the apparent optical depth when inclination dependence of total brightness is studied statistically. The importance of this division into real and apparent optical depth comes from the need for accurate inclination corrections in observed galaxies to determine the values that these would have if seen face-on. This is because close optical depth studies of individual galaxies do not necessarily provide accurate inclination correction factors. Interestingly, the conclusion that galaxies are really optically thin but behave as if they were thicker contradicts the conclusions for the sandwich model presented by \citet{disney}, which claimed that spiral galaxies could be optically thick but behave as if they were thin because of the dominance of the outer crust of the stars. This underlines the fact that the statistical approach depends on the model required to interpret the distributions of and correlations between observable quantities. Furthermore, it can be unable to detect any spherically symmetric dust distribution. For example, there is evidence for reddening in the halo components of galaxies \citep{teerikorpi}.  

\subsection{The value of C in statistical studies}

The results of our statistical study are in agreement, indirectly, with the above effects, although they cannot explaine entirely the high values of optical depth. The results also agree with the widely accepted idea that the outer regions of the disks are optically thinner than the inner parts. Thirdly, the inclination corrections for apparent magnitude and apparent surface brightness do not differ significantly from the results of \citet{bottinelli95}. A difference with earlier work is appearent in inclination corrections for the isophotal diameter, which could be resolved by more accurate observations and more precise TF relation parameters. Using the same parameter values as \citet{bottinelli95}, the same method (see Sect. 2 \& 3) yields a subsample consisting almost entirely of the galaxies in the DL subsample in Sect. 3.2 and corresponding to a slope value of $C \approx 0.20$. This has no significant effect on the value of $\tau$ but decreases the difference between the slope values.

It should also be noted that the value derived from the diameter correction agrees well with the value $C=0.235$ used in the RC2 \citep{vaucouleurs76}, whereas the result of \citet{bottinelli95} agrees fairly well with the RC3 \citep{vaucouleurs91} value of $C=0.0$ for spiral galaxies. In view of the present results, we recommend that one checks whether the choice between these values of $C$ influences the final results.

\subsection{Implications on high-z studies}

Local TF studies are being conducted increasingly more using infrared data. The progress of more local TF research has been towards infrared bands. \citet{masters06} studied the SFI++ sample for \textit{I} and \citet{masters03,masters08} the 2MASS Extended Source Catalog for \textit{J}, \textit{H}, and \textit{K} bands. There have also been TF surveys at higher redshifts \citep{weiner,chiu,kassin}. These studies aim to determine the evolution in the luminosities and kinematics of spiral galaxies. Our results provide a starting point for applying extinction corrections to these data. However, the characteristics of these samples differ considerably from ours in terms of wavelength and selection criteria. Furthermore, the isophotal radii of the more distant galaxies only reach their inner regions, because of the dimming of the surface brightness caused by the redshift. The derived inclination angle is also expected to be less accurate, which does not affect the corrections of quantities with slight inclination dependence (diameter in our study), but does cause extra scatter in quantities with stronger inclination dependence (e.g. magnitude in our study). Obviously, a statistical study of a more distant sample would be a more challenging project, which may provide results that different from ours.


\section{Conclusions}

In a statistical study of opacity, we have analyzed data from the ongoing KLUN+ survey with a method that includes TF and kinematical distance information.

   \begin{enumerate}
      \item Our analysis has found that spiral galaxy disks are quite optically thin, at least in their outer regions, with an optical depth value $\tau_{\mathrm{B}} \approx  0.1$ obtained using the diameter TF relation. This differs from the conclusion of \citet{bottinelli95}, who inferred that the outer regions were optically thick.
      \item However, using apparent magnitude and apparent surface brightness as test quantities, it appears that the entire disk behaves statistically as a thick component with an apparent optical depth value $\tau_{\mathrm{B}} > 1$. This is in approximate agreement with \citet{bottinelli95} and with the magnitude-inclination correction adopted by HyperLeda. 
      \item On the other hand, various studies of face-on galaxy disks have found optically thin disks. This apparent contradiction with statistical studies based on inclination angle may be at least partly due to light scattering, geometric distribution of the dust, and incorrect measurements of the apparent diameter. These factors are expected to have an effect on the inclination correction values, which can be resolved accurately only by larger, statistical methods.
   \end{enumerate}

\begin{acknowledgements}
We thank the anonymous referee for useful comments and suggestions. We have made use of the HyperLeda database. E. K. acknowledges the support of the Finnish Academy of Science and Letters (Vilho, Yrj\"{o} and Kalle V\"{a}is\"{a}l\"{a} Foundation).
\end{acknowledgements}


\begin{thebibliography}{}

  \bibitem[Andredakis \& van der Kruit(1992)]{andredakis} Andredakis, Y.~C., \& van der Kruit, P.~C.\ 1992, \aap, 265, 396 

  \bibitem[Baes \& Dejonghe(2001)]{baes} Baes, M., \& Dejonghe, H.\ 2001, \mnras, 326, 733 

  \bibitem[Berdyugin \& Teerikorpi(2002)]{berdyugin} Berdyugin, A., \& Teerikorpi, P.\ 2002, \aap, 384, 1050 

  \bibitem[Berlind et al.(1997)]{berlind} Berlind, A.~A., Quillen, A.~C., Pogge, R.~W., \& Sellgren, K.\ 1997, \aj, 114, 107 

  \bibitem[Bosma et al.(1992)]{bosma} Bosma, A., Byun, Y., Freeman, K.~C., \& Athanassoula, E.\ 1992, \apjl, 400, L21 

  \bibitem[Botticella et al.(2008)]{botticella} Botticella, M.~T., Riello, M., Cappellaro, E., et al.\ 2008, \aap, 479, 49 

  \bibitem[Bottinelli et al.(1986)]{bottinelli86} Bottinelli, L., Gouguenheim, L., Paturel, G., \& Teerikorpi, P.\ 1986, \aap, 156, 157 

  \bibitem[Bottinelli et al.(1995)]{bottinelli95} Bottinelli, L., Gouguenheim, L., Paturel, G., \& Teerikorpi, P.\ 1995, \aap, 296, 64 

  \bibitem[Burstein et al.(1991)]{burstein} Burstein, D., Haynes, M.~P., \& Faber, S.~M.\ 1991,\nat, 353, 515 

  \bibitem[Byun(1993)]{byun93} Byun, Y.-I.\ 1993, \pasp, 105, 993 

  \bibitem[Byun et al.(1994)]{byun94} Byun, Y.~I., Freeman, K.~C., \& Kylafis, N.~D.\ 1994, \apj, 432, 114 

  \bibitem[Chiu et al.(2007)]{chiu} Chiu, K., Bamford, S.~P., \& Bunker, A.\ 2007, \mnras, 377, 806 

  \bibitem[Davies et al.(1993)]{davies} Davies, J.~I., Phillips, S., Boyce, P.~J., \& Disney, M.~J.\ 1993, \mnras, 260, 491 

  \bibitem[Di Bartolomeo et al.(1995)]{bartolomeo} Di Bartolomeo, A., Barbaro, G., \& Perinotto, M.\ 1995, \mnras, 277, 1279 

  \bibitem[Disney et al.(1989)]{disney} Disney, M., Davies, J., \& Phillipps, S.\ 1989, \mnras, 239, 939 

  \bibitem[Domingue et al.(2000)]{domingue} Domingue, D.~L., Keel, W.~C., \& White, R.~E., III 2000, \apj, 545, 171 

  \bibitem[Driver et al.(2007)]{driver} Driver, S.~P., Popescu, C.~C., Tuffs, R.~J., et al.\ 2007, \mnras, 379, 1022 

  \bibitem[Fouqu\'{e} \& Paturel(1985)]{fouque} Fouqu\'{e}, P., \& Paturel, G.\ 1985, \aap, 150, 192 

  \bibitem[Freeman(1970)]{freeman} Freeman, K.~C.\ 1970, \apj, 160, 811 

  \bibitem[Giovanelli et al.(1994)]{giovanelli} Giovanelli, R., Haynes, M.~P., Salzer, J.~J., et al.\ 1994, \aj, 107, 2036 

  \bibitem[Gonz{\'a}lez et al.(1998)]{gonzalez} Gonz{\'a}lez, R.~A., Allen, R.~J., Dirsch, B., et al.\ 1998, \apj, 506, 152 

  \bibitem[Guillard(2004)]{guillard} Guillard, P.\ 2004, Magist\`{e}re program report, Universit\'{e} Paris Sud

  \bibitem[Hatano et al.(1997)]{hatano97} Hatano, K., Fisher, A., \& Branch, D.\ 1997, \mnras, 290, 360 

  \bibitem[Hatano et al.(1998)]{hatano98} Hatano, K., Branch, D., \& Deaton, J.\ 1998, \apj, 502, 177 

  \bibitem[Holmberg(1946)]{holmberg} Holmberg, E. 1946, Medd. Lund. Ser. II, No. 117

  \bibitem[Holwerda et al.(2005a)]{holwerda05a} Holwerda, B.~W., Gonzalez, R.~A., Allen, R.~J., \& van der Kruit, P.~C.\ 2005a, \aj, 129, 1381 

  \bibitem[Holwerda et al.(2005b)]{holwerda05b} Holwerda, B.~W., Gonzalez, R.~A., Allen, R.~J., \& van der Kruit, P.~C.\ 2005b, \aj, 129, 1396 

  \bibitem[Huizinga \& van Albada(1992)]{huizinga} Huizinga, J.~E., \& van Albada, T.~S.\ 1992, \mnras, 254, 677 

  \bibitem[Jones et al.(1996)]{jones} Jones, H., Davies, J.~I., \& Trewhella, M.\ 1996, \mnras, 283, 316

  \bibitem[Kassin et al.(2007)]{kassin} Kassin, S.~A., Weiner, B.~J., Faber, S.~M., et al.\ 2007, \apjl, 660, L35 

  \bibitem[K\"{o}ppen \& Vergely(1998)]{koppen} K\"{o}ppen, J., \& Vergely, J.-L.\ 1998, \mnras, 299, 567 

  \bibitem[Lauberts \& Valentijn(1989)]{lauberts} Lauberts, A., \& Valentijn, E.~A.\ 1989, The Surface Photometry Catalogue of the ESO-Uppsala Galaxies, European Southern Observatory, Munich

  \bibitem[Masters et al.(2003)]{masters03} Masters, K.~L., Giovanelli, R., \& Haynes, M.~P.\ 2003, \aj, 126, 158 

  \bibitem[Masters et al.(2006)]{masters06} Masters, K.~L., Springob, C.~M., Haynes, M.~P., \& Giovanelli, R.\ 2006, \apj, 653, 861 

  \bibitem[Masters et al.(2008)]{masters08} Masters, K.~L., Springob, C.~M., \& Huchra, J.~P.\ 2008, \aj, 135, 1738 

  \bibitem[Paturel \& Teerikorpi(2005)]{paturel05} Paturel, G., \& Teerikorpi, P.\ 2005, \aap, 443, 883 

  \bibitem[Paturel et al.(1997)]{paturel97} Paturel, G., Andernach, H., Bottinelli, L., et al.\ 1997, \aaps, 124, 109 

  \bibitem[Paturel et al.(2000)]{paturel00} Paturel, G., Fang, Y., Petit, C., Garnier, R., \& Rousseau, J.\ 2000, \aaps, 146, 19 

  \bibitem[Paturel et al.(2003a)]{paturel03a} Paturel, G., Petit, C., Prugniel, Ph., et al.\ 2003a, \aap, 412, 45 

  \bibitem[Paturel et al.(2003b)]{paturel03b} Paturel, G., Theureau, G., Bottinelli, L., et al.\ 2003b, \aap, 412, 57 

  \bibitem[Peletier et al.(1994)]{peletier} Peletier, R.~F., Valentijn, E.~A., Moorwood, A.~F.~M., \& Freudling, W.\ 1994, \aaps, 108, 621 

  \bibitem[Riello \& Patat(2005)]{riello} Riello, M., \& Patat, F.\ 2005, \mnras, 362, 671 

  \bibitem[Schlegel et al.(1998)]{schlegel} Schlegel, D.~J., Finkbeiner, D.~P., \& Davis, M.\ 1998, \apj, 500, 525 

  \bibitem[Shao et al.(2007)]{shao} Shao, Z., Xiao, Q., Shen, S., et al.\ 2007, \apj, 659, 1159

  \bibitem[Simien \& de Vaucouleurs(1986)]{simien} Simien, F., \& de Vaucouleurs, G.\ 1986, \apj, 302, 564

  \bibitem[Teerikorpi(2002)]{teerikorpi} Teerikorpi, P.\ 2002, \aap, 386, 865 

  \bibitem[Theureau et al.(1998)]{theureau98} Theureau, G., Bottinelli, L., Coudreau-Durand, N., et al.\ 1998, \aaps, 130, 333 

  \bibitem[Theureau et al.(2005)]{theureau05} Theureau, G., Coudreau, N., Hallet, N., et al.\ 2005, \aap, 430, 373 

  \bibitem[Theureau et al.(2007)]{theureau07} Theureau, G., Hanski, M.~O., Coudreau, N., Hallet, N., \& Martin, J.-M.\ 2007, \aap, 465, 71 

  \bibitem[Theureau et al.(2008)]{theureau08} Theureau, G., Coudreau, N., Hallet, N., \& Martin, J.\-M. 2008, \aap, in prep.

  \bibitem[Tully et al.(1998)]{tully} Tully, R.~B., Pierce, M.~J., Huang, J.-S., et al.\ 1998, \aj, 115, 2264 

  \bibitem[Valentijn(1990)]{valentijn} Valentijn, E.~A.\ 1990, \nat, 346, 153 
  
  \bibitem[Vaucouleurs et al.(1976)]{vaucouleurs76} Vaucouleurs, G. de, Vaucouleurs, A. de, \& Corwin, H. G. Jr.\ 1976, Second Reference Catalogue of Bright Galaxies, University of Texas Press, Austin (RC2)

  \bibitem[Vaucouleurs et al.(1991)]{vaucouleurs91} Vaucouleurs, G. de, Vaucouleurs, A. de, Corwin, H. G. Jr., et al.\ 1991, Third Reference Catalogue of Bright Galaxies, Springer-Verlag (RC3)

  \bibitem[Weiner et al.(2006)]{weiner} Weiner, B.~J., Willmer, C.~N.~A., Faber, S.~M., et al.\ 2006, \apj, 653, 1049 

  \bibitem[White \& Keel(1992)]{white92} White, R.~E., III \& Keel, W.~C.\ 1992, \nat, 359, 655 

  \bibitem[White et al.(2000)]{white00} White, R.~E., III, Keel, W.~C., \& Conselice, C.~J.\ 2000, \apj, 542, 761 

  \bibitem[Xilouris et al.(1997)]{xilouris97} Xilouris, E.~M., Kylafis, N.~D., Papamastorakis, J., Paleologou, E.~V., \& Haerendel, G.\ 1997, \aap, 325, 135 

  \bibitem[Xilouris et al.(1998)]{xilouris98} Xilouris, E.~M., Alton, P.~B., Davies, J.~I., et al.\ 1998, \aap, 331, 894 

  \bibitem[Xilouris et al.(1999)]{xilouris99} Xilouris, E.~M., Byun, Y.~I., Kylafis, N.~D., Paleologou, E.~V., \& Papamastorakis, J.\ 1999, \aap, 344, 868   


\end{thebibliography}
\end{document}